\documentclass[reprint,pra,superscriptaddress,twocolumn]{revtex4}
\usepackage{units}
\usepackage{amsmath}
\usepackage{amssymb}
\usepackage{graphicx}
\usepackage{bm}
\usepackage{microtype,color}

\newcommand{\be}{\begin{equation}}
\newcommand{\ee}{\end{equation}}
\newcommand{\bea}{\begin{eqnarray}}
\newcommand{\eea}{\end{eqnarray}}

\newcommand{\e}{\epsilon}

\begin{document}
\title{Rotation-induced Asymmetry of Far-field Emission from Optical Microcavities}

\author{Li Ge}
\email{li.ge@csi.cuny.edu}
\affiliation{\textls[-20]{Department of Engineering Science and Physics, College of Staten Island, CUNY, Staten Island, NY 10314, USA}}
\affiliation{\textls[-20]{The Graduate Center, CUNY, New York, NY 10016, USA}}
\author{Raktim Sarma}
\affiliation{Department of Applied Physics, Yale University, New Haven, CT 06520-8482, USA}
\author{Hui Cao}
\email{hui.cao@yale.edu}
\affiliation{Department of Applied Physics, Yale University, New Haven, CT 06520-8482, USA}

\date{\today}
\begin{abstract}
We study rotation-induced asymmetry of far-field emission from optical microcavities, based on which a new scheme of rotation detection may be developed. It is free from the ``dead zone" caused by the frequency splitting of standing-wave resonances at rest, in contrast to the Sagnac effect. A coupled-mode theory is employed to provide a quantitative explanation and guidance on the optimization of the far-field sensitivity to rotation.
We estimate that a $10^4$ enhancement of the minimal detectable rotation speed can be achieved by measuring the far-field asymmetry, instead of the Sagnac effect, in microcavities 5 microns in radius and with distinct emission directions for clockwise and counterclockwise waves.
\end{abstract}
\maketitle

Optical microcavities have found a wide range of applications from coherent light sources in integrated photonic circuits to cavity quantum electrodynamics, single-photon emitters, and biochemical sensors \cite{Chang,Vahala}. Recently they have also been proposed as a platform for rotation detection \cite{Harayama_PRA06,Harayama_OpEx07,Scheuer2}, replacing their tabletop counterparts in optical gyroscopes for reduced system size and weight \cite{Matsko,Stenberg1,Stenberg2,Scheuer1,Stenberg3,Peng,Sorrentino,Novitski}.
An optical gyroscope utilizes the Sagnac effect \cite{Sagnac,Scully,Aronowitz,Ciminelli,Terrel}, which manifests as a rotation-induced phase shift in a non-resonant structure or frequency splitting in a resonant cavity, between two counter-propagating waves. It has several advantages over its mechanical counterparts, including the absence of moving parts and system simplicity, with a high resolution typically less than 1~deg/h.

The Sagnac effect is proportional to the size of the cavity, which puts optical microcavities at a serious disadvantage when compared with macroscopic cavities. Thus a new detection scheme is needed to make optical microcavities a viable option for rotation sensing. Previous studies \cite{Scheuer2,Sarma_JOSAB2012} indicate that the quality ($Q$) factors of two counter-propagation modes also display a rotation-induced splitting, and its relative change can be much higher than that of the resonant frequencies. This enhancement however is still not large enough to compensate for the small size of microcavities, with a sensitivity still far below the Sagnac effect in macroscopic cavities.

In this Letter we propose to use the asymmetry of the far-field emission pattern of deformed microcavity lasers as a measurable signature of rotation, which shows surprisingly high sensitivity. In a perfectly circular cavity, the output directionality of a resonance remains isotropic with rotation (see Appendix \ref{appendix:circular}). Therefore, we need to employ asymmetric resonant cavities (ARCs) \cite{Nockel,Gmachl,wiersig_prl08} to obtain directional emission so that we can detect the change in output directionality by rotation. As the rotation speed increases, either the CW or CCW wave in a resonance gradually become the dominant component, and consequently the far-field intensity pattern changes appreciably if the CW and CCW waves have very different output directionality.

\begin{figure}[htbp]
\centering
\includegraphics[width=0.93\linewidth]{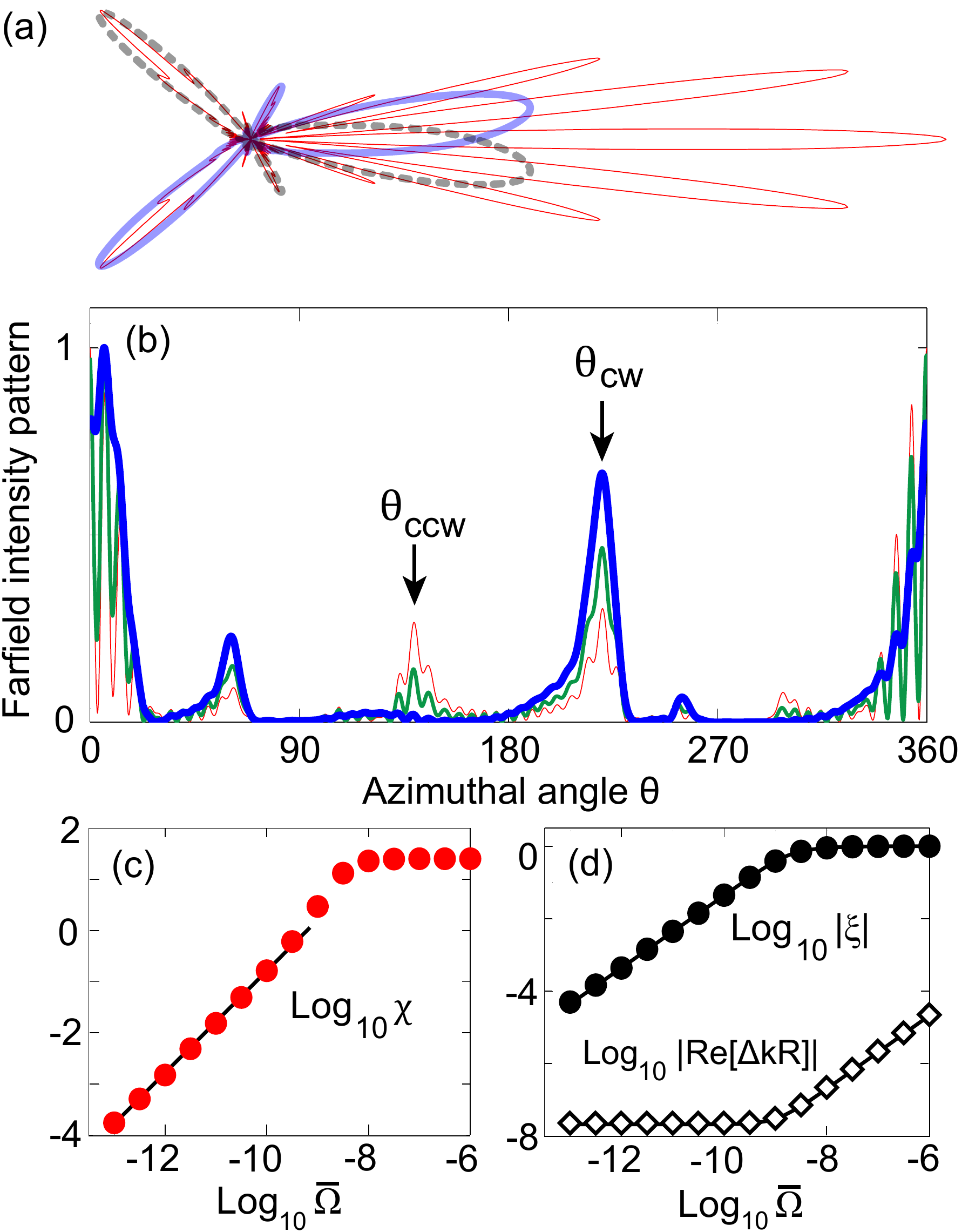}
\caption{(Color online) Evolution of the asymmetry in far-field emission from a lima\c{c}on and comparison with the Sagnac effect. The cavity deformation is $\e=0.41$ and the refractive index is $n=3$. A resonance at $k_0R\approx33.78$ and symmetric about the horizontal axis at rest is considered. (a) Far-field intensity patterns in the polar coordinates of CW (thick solid) and CCW (thick dashed) waves at rest and their superposition (thin solid) in the symmetric resonance.
(b) Far-field intensity pattern $I(\theta)$ calculated at the normalized rotation speed $\bar{\Omega} = 10^{-8} (thick line), 10^{-9} (medium line), 10^{-10}$ (thin line). The maximum intensity is normalized to unity for each curve. (c) Numerical data (dots) of far-field asymmetry $\chi$ displaying no ``dead zone" at low rotation speed. The solid line is a linear fit.
(d) Calculation by the coupled-mode theory (solid lines) agrees quantitatively with the numerical simulation of $\xi(\bar\Omega)$ (dots) - the ratio of the amplitude of $\psi^-$ and $\psi^+$ in this resonance, and the Sagnac effect - normalized frequency splitting $Re[\Delta k R]$ (diamonds). The latter has a ``dead zone'' below $\bar\Omega_c\approx1.07\times10^{-9}$.
} \label{fig:largeLimacon}
\end{figure}

One well-studied non-rotating ARC with directional emission is the lima\c{c}on cavity \cite{wiersig_prl08}, defined by $\rho(\theta) = R(1+\e\cos\theta)$ in the polar coordinates, where $R$ is the average radius and $\e$ is the deformation parameter. The main emission direction of both CW and CCW waves is in the forward direction ($\theta\approx0$) for a wide range of $\e$, but for a transverse magnetic (TM) mode the CW and CCW waves also have a significant peak in $\theta\in[120^\circ,150^\circ]$ and $[210^\circ,240^\circ]$ respectively, located symmetrically above the horizontal axis [see Fig.~\ref{fig:largeLimacon}(a)]. The CW and CCW waves couple to form non-degenerate standing-wave resonances when the cavity does not rotate. Their frequency splitting barely changes at low rotation speed until a critical value is reached \cite{Harayama_PRA06}. Such a ``dead zone'' limits the minimal speed in rotation sensing in microcavities using the Sagnac effect. On the other hand, a gradual change of the weights of the CW and CCW waves in a resonance due to rotation leads to an asymmetry of the far-field intensity pattern. We find that this asymmetry increases \textit{linearly} at low rotation speed, which is then free from the ``dead zone" that plagues the Sagnac effect.

This finding is analyzed using a coupled-mode theory with no free parameters, which agrees quantitatively with our numerical results. In addition to choosing a cavity that maximizes the difference of the output directions of the CW and CCW waves, the coupled-mode theory reveals two key quantities in optimizing the sensitivity of the far-field asymmetry, i.e. by increasing the coupling constant $g$ between the standing-wave resonances and reducing the resonance splitting $\Delta k_0$ at rest. The mode coupling constant is limited by the size of the microcavity, but the resonance splitting can be reduced by employing a cavity of high symmetry groups, such as the $D_3$ cavity given by $\rho(\theta) = R(1+\e\cos3\theta)$ \cite{Harayama_OpEx07}, and its value is only limited by surface roughness of the cavity.
We estimate that a $10^4$ enhancement of the minimal detectable speed can be achieved by measuring the far-field asymmetry instead of the Sagnac effect in $D_3$ microcavities of a $5\,\mu{m}$ radius.

Below we focus on the TM polarized resonances in two-dimensional (2D) microcavities without loss of generality. Their electric field is in the cavity plane, and their magnetic field, represented by $\psi(\vec{r})$, is perpendicular to the cavity plane. To the leading order of the rotation speed $\Omega$, the resonances are determined by the modified Helmholtz equation \cite{Harayama_PRA06}
\be
\left[\nabla^2 + n(\vec{r})^2k^2 + 2ik\frac{\Omega}{c}\frac{\partial}{\partial \theta}\right]\psi(\vec{r}) = 0,\label{eq:0}
\ee
where $n(\vec{r})$ is the refractive index, $k$ is the complex frequency of mode $\psi(\vec{r})$ in the unit of inverse length, and $\theta$ is the azimuthal angle.
We have assumed that the angular velocity is perpendicular to the cavity plane and that $\Omega>0$ indicates a CCW rotation.

To find the resonances in a rotating ARC and their far-field intensity patterns, one can use the finite-difference time-domain method adapted to the rotating frame \cite{Sarma_JOSAB2012}.
Here we employ a more effective and grid-free approach, the scattering matrix method. It applies generally to a concave cavity with a uniform refractive index and a smooth boundary deviation $\delta\rho(\theta)$ from a circle, satisfying the Rayleigh criterion $|\delta\rho(\theta)|\ll R$.
In this approach the wave function of a resonance is decomposed in the angular momentum basis, i.e. $\psi(\vec{r})=\sum_m A_m(r)e^{im\theta}$, where
\begin{gather}
\hspace{-0.5mm}
A_m(r)=\begin{cases}
\alpha_mH^+_m(\bar{k}_mr) + \beta_mH^-_m(\bar{k}_mr), & r<\rho(\theta), \\
\gamma_mH^+_m(\tilde{k}_mr), & r>\rho(\theta),
\end{cases}\label{eq:expansion}
\end{gather}
and $H^{\pm}$ are the Hankel functions of the first (outgoing) and second (incoming) kind.
Compared with the non-rotating cavities \cite{Narimanov_PRL1999,Smatrix}, the difference lies in the $m$-dependent frequencies $\bar{k}_m \equiv [(nk)^2-2mk\bar{\Omega}/R]^\frac{1}{2}$ and
$\tilde{k}_m \equiv [k^2-2mk\bar{\Omega}/R]^\frac{1}{2}$,
where $\bar{\Omega}\equiv R\Omega/c$ is the dimensionless rotation speed. The details of the scattering matrix formulation are given in Appendix \ref{appendix:smatrix}.

We first apply the method above to a lima\c{c}on cavity of $\e=0.41$, whose circumference is about 100 times the wavelength inside the cavity. The angular momentum distributions of $\alpha_m$ (and $\beta_m$) in the corresponding resonances are centered about $|m_0|\approx100$.
Due to the mirror symmetry of the cavity, the wave functions of the resonances at rest are standing waves, either symmetric ($\psi^+$) or anti-symmetric ($\psi^-$) about the horizontal axis, resulting in a symmetric intensity pattern. Next we consider as an example the symmetric resonance $\psi^+$, which is a whispering-gallery type mode with $|m_0|=101$. Its resonant frequency is about $k_mR\simeq33.78$, corresponding to a vacuum wavelength $\lambda\approx930$ nm if the average radius of the cavity is $R=5\,\mu{m}$. Its emission results from the chaotic diffusion along the unstable manifolds \cite{SI,wiersig_prl08,Schwefel}, and the two secondary peaks, located near $\theta_{ccw}=138^\circ, \theta_{cw}=222^\circ$, are from the CCW and CW waves, respectively [Fig.~\ref{fig:largeLimacon}(a)]. The intensity of the main peak at $\theta=0$ is almost four times as large, due to constructive interference of the CW and CCW peaks nearby.
As the cavity rotates, the initial balance between the CW and CCW waves is broken, similar to the finding in closed billiards \cite{Harayama_PRA06}. In this case the weights of the CW waves become larger than their CCW counterparts. As a result, the intensity peak at $\theta_{cw}$ grows with respect to the one at $\theta_{ccw}$, as well as to the main one at $\theta=0$  [see Fig.~\ref{fig:largeLimacon}(b)]. The opposite takes place in the corresponding anti-symmetric resonance $\psi^-$ of the same $|m_0|$, with the CCW waves becoming the prevailing component and the intensity peak at $\theta_{ccw}$ increased. Our discussion below is based on the far-field intensity pattern of $\psi^+$.

Utilizing the far-field difference of the CW and CCW waves, the asymmetry of the far-field intensity pattern $I(\theta)$ in the originally symmetric resonance $\psi^+$ can be measured by the ratio of peak intensity at $\theta_{cw}$ to that at $\theta_{ccw}$
\be
\chi = \frac{\int_{\,\theta_{cw}-{\Delta\theta}/{2}}^{\theta_{cw}+{\Delta\theta}/{2}} \;I(\theta)d\theta}{\int_{\,\theta_{ccw}-{\Delta\theta}/{2}}^{\theta_{ccw}+{\Delta\theta}/{2}} I(\theta)d\theta}-1,
\ee
where $\Delta\theta$ is the detection range of each peak and taken to be $15^\circ$.
It does not display a ``dead zone" at low rotation speed: $\chi$ increases linearly with $\Omega$ [Fig.~\ref{fig:largeLimacon}(c)], which is in stark contrast with the Sagnac effect; the latter barely changes until the rotation speed is higher than a critical value $\bar\Omega_c\equiv R\Omega_c/c\sim10^{-9}$ [Fig.~\ref{fig:largeLimacon}(d)]. If we assume that the far-field asymmetry can be measured experimentally down to $\sim10^{-4}$, then the minimal detectable speed is about $\bar\Omega\sim10^{-13}$, which is $10^4$ times lower that the onset of the Sagnac effect at $\bar\Omega_c$.
$\chi$ starts to saturate only at very high speed in the log-log plot, when $\bar\Omega>\bar\Omega_c$ and the CW waves dominate over the CCW waves. This does not imply that $\Omega_c$ is the upper operation limit of our approach, as the change of the far-field asymmetry in the linear scale continues to be appreciable in the whole range of rotation speed shown in Fig.~\ref{fig:largeLimacon}(c).

To understand why the far-field asymmetry does not display a ``dead zone" at low rotation speed while the Sagnac effect does, we first note that the increase of the asymmetry can be viewed as a result of the mixing of the anti-symmetric resonance $\psi^-$ with the symmetric one. Below we employ a coupled-mode theory to capture this behavior, which is similar to that given in Refs.~\cite{Harayama_PRA06,Harayama_OpEx07,Sunada_PRE08}. It applies both in and beyond the ``dead zone'' and takes into account the phase of the coupling constant. Since $\psi^+$ and $\psi^-$ are quasi-degenerate, their mutual coupling is much stronger than that with any resonance farther away in frequency \cite{BD1,BD2}. Therefore, it is a good approximation to write their wave function as $\psi(\Omega) \approx a^+(\Omega)\psi^+ + a^-(\Omega)\psi^-$ when discussing how they evolve towards the CW and CCW resonances with rotation. By substituting $\psi_m(\Omega)$ in Eq.~(\ref{eq:0}) by this expansion and solving the resulting matrix equation for $(a^+,a^-)$ (see Appendix \ref{appendix:CMT}), we find two solutions that correspond to the CW- and CCW-prevailing resonances at $\Omega\neq0$.
The mixing ratio $\xi(\Omega) \equiv {a^-}/{a^+}$ in the initial $\psi^+$ resonance is given by
\begin{align}
\xi(\Omega)^2
\approx \frac{D -\sqrt{D^2+\left( {2g^2}/{c^2} \right) \Sigma\Omega^2}}{D +\sqrt{D^2+ \left( {2g^2}/{c^2} \right) \Sigma\Omega^2}},
\label{eq:mixing}
\end{align}
where $D\equiv {k_0^+}^2-{k_0^-}^2$, $\Sigma\equiv {k_0^+}^2+{k_0^-}^2$, and $k_0^\pm$ are the two resonant frequencies at rest.  $g\equiv2\sqrt{-G_{\!\scriptscriptstyle -+}G_{\!\scriptscriptstyle +-}}/n^2$ is the dimensionless coupling constant between the standing-wave resonances $\psi^-$ and $\psi^+$, where $G_{\!\scriptscriptstyle +-} = \int_\text{cavity} \psi^+ {\partial_\theta\psi^-}\,d\vec{r}$ and $G_{\!\scriptscriptstyle -+}$ is defined similarly.
We emphasize that $g$ should be differentiated from the coupling of CW and CCW waves in the non-rotating cavity.
It can be shown that $G_{\!\scriptscriptstyle +-}\approx-G_{\!\scriptscriptstyle -+}$ in a cavity slightly deformed from a circular disk, and $g$ is approximately real and positive as a result.
The mixing ratio in the initial $\psi^-$ resonance is given by the inverse of Eq.~(\ref{eq:mixing}).

Due to the resonance splitting at rest, $D\neq0$ in a lima\c{c}on cavity, $\psi^+$ and $\psi^-$ gradually evolve towards the CW and CCW resonances as $\Omega$ increases, and $\xi(\Omega)\rightarrow -1$. It is important to note that in this process the $\Omega$-dependence of $\xi$ is {\it linear} inside the ``dead zone,'' as can be seen from
\be
\xi(\Omega)\approx \pm i\frac{\Omega}{\sqrt{2}\Omega_c}, \label{eq:linear}
\ee
for $\Omega\ll \Omega_c\equiv c|k_0^+-k_0^-|/g$. This leads to the linear dependence of $\chi$ on $\Omega$ we have seen in Fig.~\ref{fig:largeLimacon}(c).
On the other hand, the difference of the two resonances is given by
\be
\Delta k (\Omega) = \left[(\Delta k_0)^2 + \left(\frac{g}{c}\Omega\right)^2\right]^{\frac{1}{2}}, \label{eq:Deltak}
\ee
where $\Delta k_0=k_0^+-k_0^-$.
It shows a ``dead zone" for $\Omega \lesssim \Omega_c$: the leading $\Omega$-dependence of $\Delta k(\Omega)$ is quadratic at low rotation speed, and the sensitivity is reduced by a factor of $\Omega/2\Omega_c$ when compared with the initially degenerate case ($\Delta k_0=0$). Far beyond the ``dead zone,'' $\Delta k(\Omega)$ approaches its asymptote $g\Omega/c$, which is the same as the initially degenerate case.

To check the validity of the coupled-mode theory, we compare the values of $\xi(\Omega)$ and $\Delta k(\Omega)$ given by Eqs.~(\ref{eq:mixing}) and (\ref{eq:Deltak}) with the numerical result from the scattering matrix approach. Good agreement is found for the resonances with $|m_0|=101$ discussed above, with no free parameters [Fig.~\ref{fig:largeLimacon}(d)]; the only inputs of the coupled-mode theory are $g=21.45 - 0.004i$ and $\Delta k_0R=(2.29+0.90i)\times10^{-8}$ obtained from the scattering matrix method, or equivalently, $\bar\Omega_c=1.07\times10^{-9}$.

\begin{figure}[b]
\centering
\includegraphics[width=0.9\linewidth]{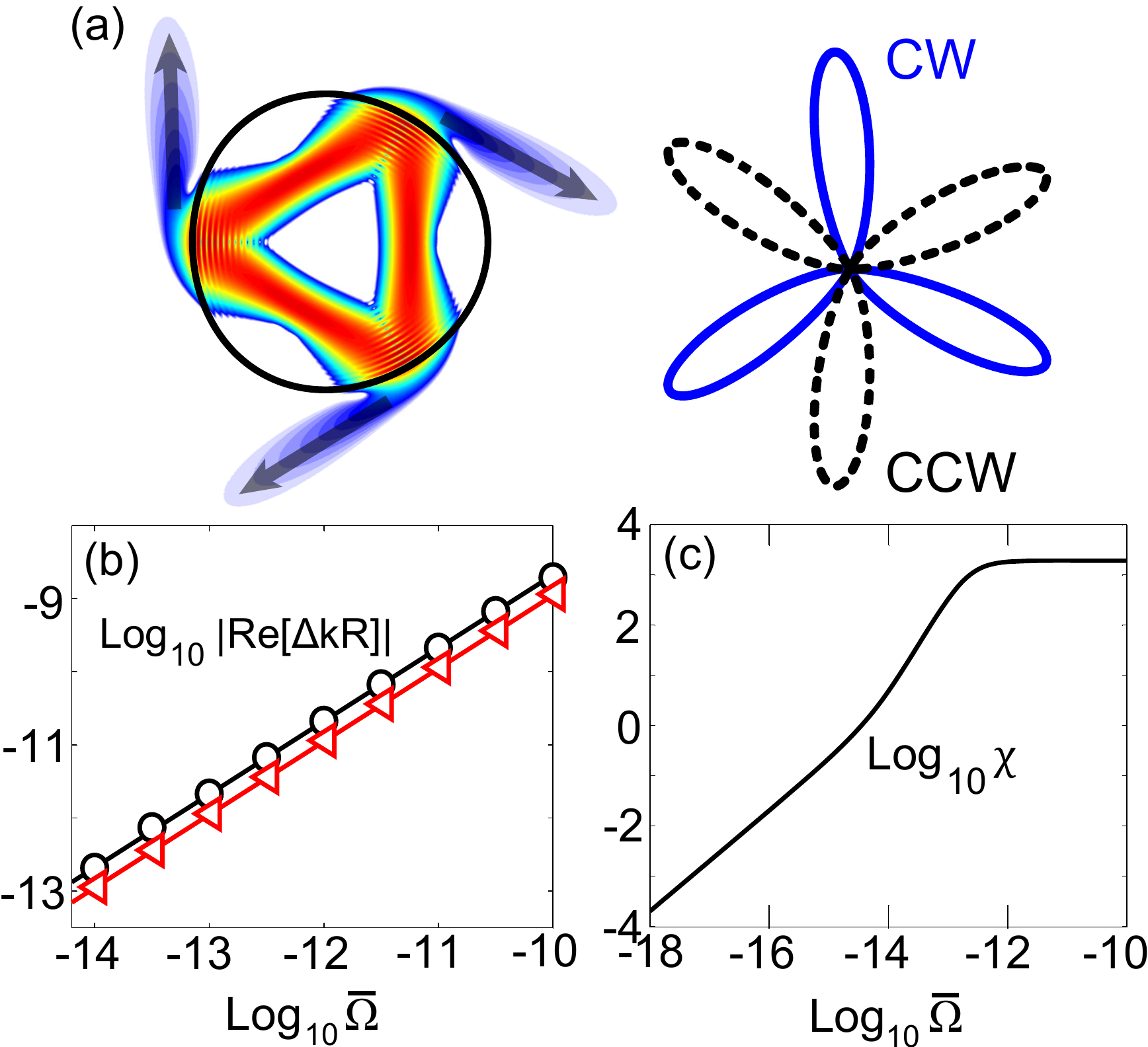}
\caption{(Color online) High rotation sensitivity of emission from a $D_3$ cavity of $\e=0.025$ and refractive index $n=3$.
(a) Left: False-color plot of the near-field pattern for the CW waves in a triangular resonance $\psi^+$ at $kR\approx33.45$ in the logarithmic scale. Right: Distinct far-field intensity patterns in the polar coordinates of CW (thick solid) and CCW (thick dashed) waves.
(b) Sagnac frequency splitting of the triangular resonance in (a) and its anti-symmetric partner $\psi^-$ calculated by the scattering matrix method (triangles). The same is shown for a pair of whispering-gallery resonances at $kR\approx33.80$ (dots). The two linear lines are plotted using Eq.~(\ref{eq:Deltak}) with the coupling constant $g=11.06, \, 20.62$ obtained from the scattering matrix method.
(c) Estimation of the far-field asymmetry for the CW-prevailing triangular resonance, using $\xi(\theta)$ from the coupled-mode theory and the CW and CCW wave functions from the scattering matrix method. $\bar\Omega_c=10^{-14}$ is assumed, and $\theta_{cw}=94^\circ$, $\theta_{ccw}=266^\circ$, and $\Delta\theta=15^\circ$ are used in calculating $\chi$.}
\label{fig:D3}
\end{figure}

Next we analyze how to optimize the far-field asymmetry for rotation sensitivity. First of all, the CW and CCW waves must have very different emission directionality, so that the far-field asymmetry changes significantly as a function of the rotation speed. Once this condition is satisfied, we note that the linear increase of the far-field asymmetry at low rotation speed is due to the relation (\ref{eq:linear}). Thus we need to reduce $\Omega_c$, or equivalently, reduce the resonance splitting $|\Delta k_0|$ at rest and increase the coupling constant $g$ between $\psi^+$ and $\psi^-$. From the definition of $G_{\!\scriptscriptstyle +-},G_{\!\scriptscriptstyle -+}$, we know that they are roughly proportional to $|m_0|$, which is approximately $nk_0^{\pm}R$ for whispering-gallery resonances if $|m_0|\gg1$ \cite{JensThesis}. Therefore, $g$ is proportional to $k_0R$ to a good approximation, which is expected as it is the coefficient in the asymptote of the Sagnac effect. Thus it scales linear with the cavity size and the operational frequency. On the other hand, $\Delta k_0$ can be reduced by using microcavities of higher symmetry groups, such as the $D_3$ cavity with $\rho(\theta)=R(1+\e\cos3\theta)$ \cite{Harayama_OpEx07}. Idealistically $\Delta k_0$ can be entirely eliminated for resonances in the $D_3$ cavity, if they are not simultaneously the eigenfunctions of the parity about the horizontal axis and $2\pi/3$ rotation, or equivalently, if their angular momenta are not integer times of 3.
In practice, there is always inherent surface roughness introduced unintentionally during the fabrication process, which breaks the exact $D_3$ symmetry and lift the degeneracy of the CW and CCW waves at rest slightly. In the scattering matrix approach, the limited precision in carrying out the boundary integral (see Appendix \ref{appendix:smatrix}) can also be regarded as one type of surface roughness. Nevertheless, for the idealistically degenerate high-$Q$ resonances near $nk_0^\pm R\sim100$ in a $D_3$ cavity of $\e=0.025$ and $n=3$, we find that they all have a $|\Delta k_0R|$ below our numerical accuracy ($\sim10^{-13}$). Furthermore, from the linear behavior of $\Delta k(\Omega)$ shown in Fig.~\ref{fig:D3}(b), we know that for $\bar\Omega > 10^{-14}$ the frequency splitting is already in the asymptotic region of the Sagnac effect. Thus the critical speed $\bar\Omega_c$ is below $10^{-14}$.

If we take this upper bound as a practically realizable value of $\bar\Omega_c$, (or equivalently $\Delta k_0R\sim10^{-15}$ given $g\sim10$),
Eq.~(\ref{eq:linear}) then shows that the sensitivity of the mixing ratio $\xi(\Omega)$ in this $D_3$ cavity is about $10^5$ times higher than the lima\c{c}on cavity of the same average radius studied above. We note that the CW and CCW waves in the triangular resonance shown in Fig.~\ref{fig:D3}(a) have distinct far-field emission directions. By constructing the far-field asymmetry using $\xi(\Omega)$ given by the coupled-mode theory, the coupling constant $g=11.06$ from the scattering matrix method, and assuming  $\bar\Omega_c=10^{-14}$, we find that a rotation sensitivity about $\bar\Omega\sim10^{-18}$ ($\Omega\sim 10$~deg/h)
may be achieved when the far-field asymmetry can be measured down to $\sim10^{-4}$ experimentally [Fig.~\ref{fig:D3}(c)], in a microcavity just $5\,\mu{m}$ in radius.
We note that this minimal speed is much lower than $\bar\Omega_c$ (by a factor of $10^4$), the onset speed of the Sagnac effect in the same microcavity. The whispering-gallery resonances shown in Fig.~\ref{fig:D3}(b) have directional emission and a similar sensitivity in this estimation (see Appendix \ref{appendix:nearfield}).

In summary, we have proposed to measure the far-field asymmetry as a promising approach for rotation sensing in deformed microcavities, which is free from the ``dead zone" that plagues the Sagnac effect. The minimal detectable rotation speed is estimated to be $10^4$ smaller than the onset speed of the Sagnac effect in the same microcavity, which holds in both the lima\c{c}on and $D_3$ cavities discussed above.
The sensitivity can be further increased by using a larger cavity, to which the critical speed $\Omega_c$ is inversely proportional via the coupling constant $g$. As mentioned previously, we have based our estimates on the far-field asymmetry of the $\psi^+$ resonance. For the initially anti-symmetric resonance $\psi^-$, its far-field intensity changes in opposite to that of $\psi^+$ with rotation and thus counteracts the change of the total far-field asymmetry if both resonances are excited. However, the cancelation is only partial unless they have equal intensities and they are phase incoherent or their relative phase is integer times of $\pi/4$ at one detection direction. We believe that nonlinear effects such as spatial hole burning in a homogeneously broaden gain medium will lead to different lasing intensities of these resonances, which will only lead to a fractional reduction of the far-field asymmetry and the rotation sensitivity estimated above. Detailed study of the nonlinear effects will be included in a further work.

We thank Takahisa Harayama and Jan Wiersig for helpful discussions. L.G. acknowledges PSC-CUNY 45 Research Award. R.S. and H.C. acknowledges NSF support under Grant No. ECCS-1128542.

\appendix

\section{Resonances in a rotating circular cavity}
\label{appendix:circular}
In a circular microcavity, a pair of CW and CCW modes, $\psi_m(\vec{r})\propto e^{\mp i|m|\theta}$ ($m$: angular momentum number), are degenerate when the cavity does not rotate. As a result, any mixture of them is still an eigenstate of the system, such as the standing waves proportional to $\sin(m\theta),\cos(m\theta)$.
At a nonzero rotation speed $\Omega$ however, Eq.~(1) in the main text permits only CW or CCW resonances with different frequencies (the same is true in cavities of high symmetry groups, such as the one with $D_3$ symmetry \cite{Harayama_OpEx07}). They can be found by solving the continuity equation of $\psi_m(\vec{r})$ and its radial derivative at $r=R$, i.e.
\be
\bar{k}_m\frac{J'_m(\bar{k}_mR)}{J_m(\bar{k}_mR)} = \tilde{k}_m\frac{{H^+_m}'(\tilde{k}_mR)}{H^+_m(\tilde{k}_mR)},\label{eq:BD}
\ee
in which $\bar{k}_m \equiv [(nk)^2-2mk\bar{\Omega}/R]^\frac{1}{2}$,
% = (nkR-m\bar{\Omega}/n) + O(\bar{\Omega}^2)
$\tilde{k}_m \equiv [k^2-2mk\bar{\Omega}/R]^\frac{1}{2}$,
% =  kR-m\bar{\Omega} + O(\bar{\Omega}^2)$
and $\bar{\Omega}\equiv R\Omega/c$ is the dimensionless rotation speed defined in the main text. We note that the resulting $k_m\neq k_{-m}$, CW and CCW modes do not mix even though the angular momentum is still a good quantum number; the mixing of the CW and CCW waves of the same $|m|$ found in Ref.~\cite{Sarma_JOSAB2012} is not caused by rotation but rather the excitation method in the finite-difference-time-domain method.

\section{Scattering matrix method for rotating ARCs}
\label{appendix:smatrix}
The (internal) scattering matrix $\mathcal{S}(k)$ used in the main text maps the internal incident waves on the cavity boundary ($\alpha_m$ in Eq.~(2) of the main text) to the scattered waves inside ($\beta_m$).
$\mathcal{S}(k)$ is found by solving the continuity conditions of the TM wave function and its radial derivative at the cavity boundary, which can be put into the following matrix form
\begin{align}
\mathcal{\overline{H}}^+|\alpha\rangle + \mathcal{\overline{H}}^-|\beta\rangle &= \mathcal{\tilde{H}}^+|\gamma\rangle, \label{eq:S1}\\
\mathcal{\overline{D}}^+|\alpha\rangle + \mathcal{\overline{D}}^-|\beta\rangle &= \mathcal{\widetilde{D}}^+|\gamma\rangle, \label{eq:S2}
\end{align}
where
\begin{align}
\mathcal{\overline{H}}^\pm_{lm} &= \int_0^{2\pi} H^\pm_m[\bar{k}_m\rho(\theta)] e^{i(m-l)\theta} d\theta, \\ \mathcal{\overline{D}}^\pm_{lm} &= \int_0^{2\pi} \bar{k}_m{H^\pm_m}'[\bar{k}_m\rho(\theta)] e^{i(m-l)\theta} d\theta,\label{eq:Dp}
\end{align}
and $\mathcal{\tilde{H}}^+, \, \mathcal{\widetilde{D}}^+$ are defined similarly with $\bar{k}_m$ substituted by $\tilde{k}_m$. The apostrophe in Eq.~(\ref{eq:Dp}) represents the derivative of the Hankel functions. By eliminating $\gamma_m$ from Eqs.~(\ref{eq:S1}) and (\ref{eq:S2}), a matrix equation can be found in the form $\mathcal{S}(k)|\alpha\rangle = |\beta\rangle$. Instead of solving this equation as a set of inhomogeneous linear equations, we solve it as an eigenvalue problem, $\mathcal{S}(k)|\alpha\rangle = e^{i\phi}|\alpha\rangle$ \cite{Narimanov_PRL1999,Smatrix}, in which the \textit{complex} quantity $\phi$ is defined in such a way that $|\beta\rangle= e^{i\phi}|\alpha\rangle$. The resonances $k$ are then determined by the condition $\phi=0$, so that $|\alpha\rangle=|\beta\rangle$ and the wave function inside the cavity is given by the superposition of Bessel functions, i.e. $J_m = (H_m^+ + H_m^-)/2$, which has a finite amplitude at the origin $\vec{r}=0$. All other values of $\phi$ lead to unphysical states with an infinite amplitude at $\vec{r}=0$, because $|H^\pm_m(r\rightarrow 0)|\rightarrow\infty$.

\section{Coupled-mode theory}
\label{appendix:CMT}
The coupled-mode theory for the Sagnac effect in a closed billiard was given in Refs.~\cite{Harayama_PRA06,Harayama_OpEx07}, which applies only when the rotation speed is beyond the ``dead zone." In Ref.~\cite{Sunada_PRE08} the authors extended it into the ``dead zone" but only for a ring laser. The coupled-mode theory presented in the main text shows that a similar approach can be applied to open cavities, both in and beyond the ``dead zone,'' and takes into account the phase of the coupling constant.

By substituting $\psi(\Omega) \approx a^+(\Omega)\psi^+ + a^-(\Omega)\psi^-$ to Eq.~(1) of the main text, we find that the coefficients satisfy the following equation,
\be
\left(
\begin{array}{c c}
k^2-{k_{0}^+}^2 & \frac{2ik\Omega}{cn^2}G_{\!\scriptscriptstyle +-} \\
\frac{2ik\Omega}{cn^2}G_{\!\scriptscriptstyle -+} & k^2-{k_{0}^-}^2
\end{array}\right)
\begin{pmatrix}
a^+ \\ a^-
\end{pmatrix}
=0,\label{eq:modecoupling}
\ee
where ${k_{0}^+},{k_{0}^-}$ are the resonant frequencies at rest,
$G_{\!\scriptscriptstyle +-} = \int_\text{cavity} \psi^+ {\partial_\theta\psi^-}\,d\vec{r}$, and $G_{\!\scriptscriptstyle -+}$ is defined similarly. We note that $G_{\!\scriptscriptstyle ++}$ and $G_{\!\scriptscriptstyle --}$, which would have appeared on the diagonal of the coupling matrix in Eq.~(\ref{eq:modecoupling}), vanish because their integrands are odd functions with respect to the horizontal axis. Likewise, $\int_\text{cavity} \psi^+\psi^-\,d\vec{r}$ vanishes even though resonances of an open cavity are not orthogonal or biorthogonal in general. We have used the normalization $\int_\text{cav} (\psi^{\pm})^2\,d\vec{r}=1$.

By solving Eq.~(\ref{eq:modecoupling}), we find the expression for the coupling ratio $\xi(\Omega)$ and complex resonance splitting $\Delta k(\Omega)$ given in the main text. 

\section{Near-field patterns}
\label{appendix:nearfield}
\begin{figure}[htbp]
\centering
\includegraphics[width=0.9\linewidth]{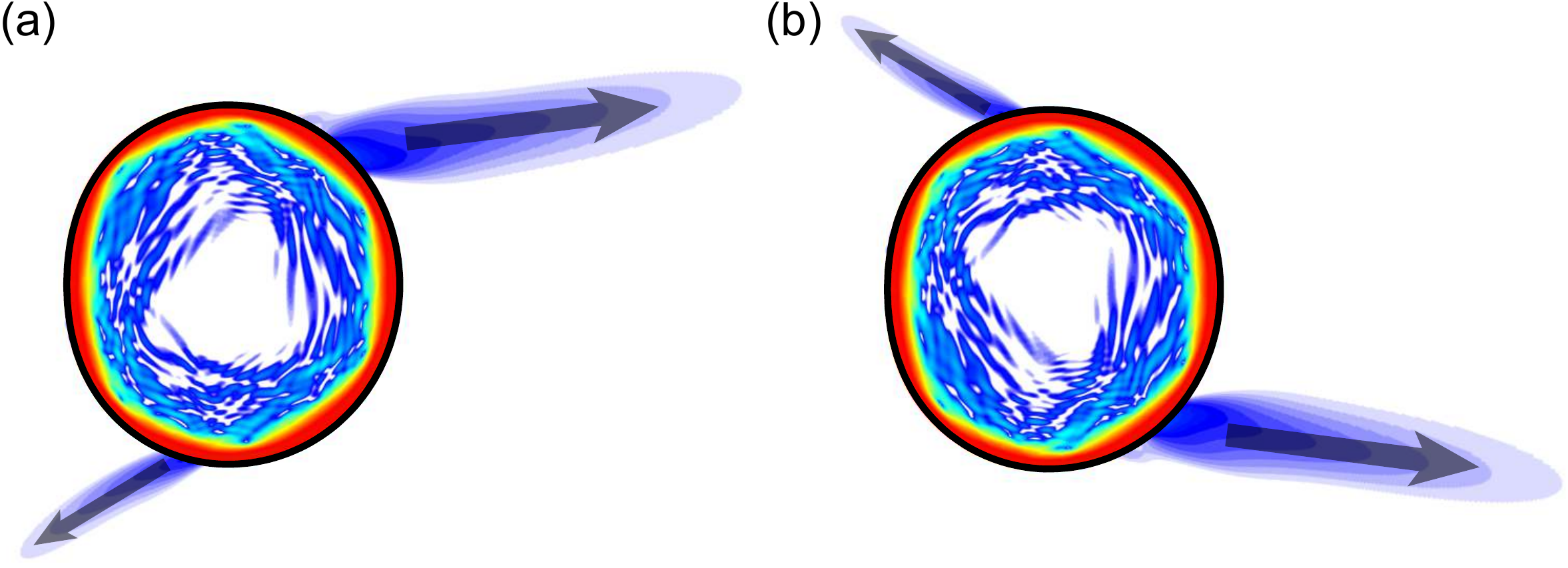}
\caption{False-color plot in the logarithmic scale showing the near-field intensity patterns of (a) the CW waves and (b) the CCW waves of the symmetric whispering-gallery resonance $\psi^+$ at $\Omega=0$ in Fig.~1(a) of the main text. } \label{fig:nearfield_limacon}
\end{figure}

\begin{figure}[t]
\centering
\includegraphics[width=0.9\linewidth]{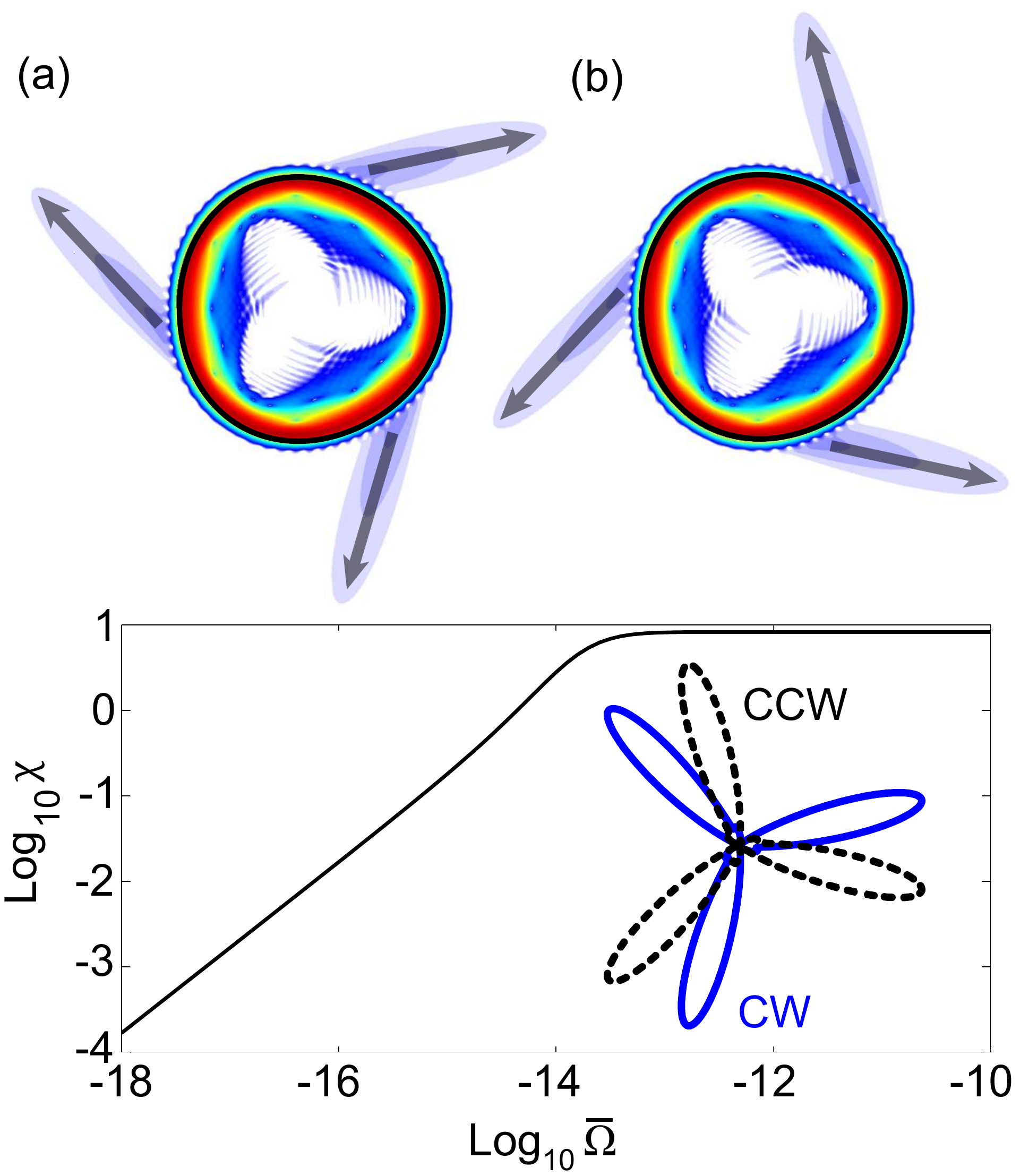}
\caption{False-color plot in the logarithmic scale showing the near-field intensity patterns of (a) the CW waves and (b) the CCW waves of a symmetric whispering-gallery resonance $\psi^+$ at $kR\approx33.80$ in the $D_3$ cavity discussed in the main text at $\Omega=0$. (c) shows the estimate of the far-field asymmetry for the $\psi^+$ resonance. Inset: Far-field intensity patterns of the CW (solid) and CCW (dashed) waves in $\psi^+$ at rest in the polar coordinates.} \label{fig:nearfield_D3}
\end{figure}

In Fig.~1(a) of the main text we have shown the far-field patterns of the CW and CCW waves of a symmetric whispering-gallery resonance $\psi^+$ in a lima\c{c}on cavity at $\Omega=0$. Their near-field patterns are shown in Fig.~\ref{fig:nearfield_limacon}, which are mirror images of each other about the horizontal axis. As the rotation speed becomes larger than the critical speed $\Omega_c$, this resonance and its anti-symmetric partner approach these CW and CCW patterns, respectively. Their patterns inside the cavity are enhanced in the plot by using the logarithmic scale, which show the chaotic diffusion of the waves that leads to the directional emission.

In the discussion of the $D_3$ cavity in the main text, we have mentioned that the whispering-gallery modes exist along side the triangular modes, and they have directional emission as well. This can be seen from Fig.~\ref{fig:nearfield_D3}(a) and (b), which show the near-field patterns of the CW and CCW waves of a symmetric whispering-gallery resonance $\psi^+$. The frequency difference of this resonance and its pairing anti-symmetric resonance is shown in Fig.~2(b) of the main text. Fig.~\ref{fig:nearfield_D3}(c) shows the estimate of the far-field asymmetry for the $\psi^+$ resonance, using $\xi(\theta)$ from the coupled-mode theory and the CW and CCW wave functions from the scattering matrix method. $\bar\Omega_c=10^{-14}$ is assumed, and $\theta_{cw}=256^\circ$, $\theta_{ccw}=106^\circ$, and $\Delta\theta=15^\circ$ are used in calculating $\chi$. It has similar sensitivity to the triangular mode shown in Fig.~\ref{fig:nearfield_limacon} and discussed in Fig.~2(c) of the main text.

\bibliographystyle{apsrev}

\end{document}